\documentclass[10pt,a4paper,final]{article}
\usepackage[utf8]{inputenc}
\usepackage{tikz}
\usepackage{braket}
\usetikzlibrary{quantikz}
\usepackage{geometry}
\newgeometry{vmargin={25mm}, hmargin={25mm,25mm}} 
\usepackage{comment}
\usepackage{listings}
\usepackage{algorithm}
\usepackage{algpseudocode}
\title{Introducing $AQCtensor$: A Tensor Network based approach to Approximate Quantum Compiling}
\title{Approximate Quantum Compiling for Quantum Simulation: A Tensor Network based approach}
\author{}
\date{}

\newcommand{\beq}{\begin{equation}}
\newcommand{\eeq}{\end{equation}}
\newcommand{\no}{\noindent}
\newcommand{\cnbl}{$CNOT$ block\hspace{0.5em}}
\newcommand{\cnbls}{$CNOT$ blocks\hspace{0.5em}}

\begin{document}

\maketitle

\begin{center}
{\large Niall F. Robertson$^{1}$, Albert Akhriev$^{1}$, Jiri Vala$^{2,3}$ and Sergiy Zhuk$^{1}$}
\end{center}

\vspace{0.5cm}

{\no \sl\small $^1$ IBM Quantum, IBM Research Europe - Dublin, IBM Technology Campus, Dublin 15, Ireland\\}
{\sl\small $^2$ Maynooth University, Maynooth, Ireland\\}
{\sl\small $^3$ Tyndall National Institute, Cork, Ireland}

\begin{abstract}
We introduce $AQCtensor$, a novel algorithm to produce short-depth quantum circuits from Matrix Product States (MPS). Our approach is specifically tailored to the preparation of quantum states generated from the time evolution of quantum many-body Hamiltonians. This tailored approach has two clear advantages over previous algorithms that were designed to map a generic MPS to a quantum circuit. First, we optimize all parameters of a parametric circuit at once using Approximate Quantum Compiling (AQC) - this is to be contrasted with other approaches based on locally optimizing a subset of circuit parameters and ``sweeping" across the system. We introduce an optimization scheme to avoid the so-called ``orthogonality catastrophe" - i.e. the fact that the fidelity of two arbitrary quantum states decays exponentially with the number of qubits - that would otherwise render a global optimization of the circuit impractical. Second, the depth of our parametric circuit is constant in the number of qubits for a fixed simulation time and fixed error tolerance. This is to be contrasted with the linear circuit Ansatz used in generic algorithms whose depth scales linearly in the number of qubits. For simulation problems on 100 qubits, we show that $AQCtensor$ thus achieves at least an order of magnitude reduction in the depth of the resulting optimized circuit, as compared with the best generic MPS to quantum circuit algorithms. We demonstrate our approach on simulation problems on Heisenberg-like Hamiltonians on up to 100 qubits and find optimized quantum circuits that have signficantly reduced depth as compared to standard Trotterized circuits.
\end{abstract}

\section{Introduction}
Tensor Networks are powerful tools that can be used to tackle a range of problems in various fields including chemistry \cite{wagner2014kohn, stoudenmire2012one}, machine learning \cite{stoudenmire2016supervised, vieijra2022generative} and quantum physics \cite{orus2019tensor, ran2020tensor}. In recent years, there have been several proposals to tackle problems in these fields by combining tensor networks with quantum computing by e.g. finding an approximate solution to a ground state problem using Matrix Product States (MPS) and then preparing the corresponding quantum state on a quantum computer for further optimization \cite{rudolph2023synergistic}. Similar strategies for quantum machine learning have also been proposed \cite{huggins2019towards}. These schemes thus motivate the design of efficient algorithms that decompose a generic Matrix Product State into a quantum circuit \cite{rudolph2023decomposition, ran2020encoding, lin2021real, dov2022approximate} - the advantages and limitations of the known approaches were discussed in detail in \cite{rudolph2023decomposition}. For example, the MPS to quantum circuit algorithm described in \cite{lin2021real} requires the compilation of a unitary matrix into $1$ and $2$-qubit gates - while the number of qubits that this unitary acts on is logarithmic in the bond dimension $\chi$ of the MPS, the number of gates required to compile this unitary is exponential in the number of qubits \cite{madden2022best} in general and quadratic in $\chi$. It was also discussed in \cite{rudolph2023decomposition} that the performance of the approach described in \cite{ran2020encoding} did not improve in a meaningful way when more layers were added to the Ansatz, thus restricting the algorithm to apply to Matrix Product States with very low bond dimensions. It was finally concluded in \cite{rudolph2023decomposition} that the approach with the best performance was a novel combination of the variational algorithm proposed in \cite{dov2022approximate} with the approach of \cite{ran2020encoding}.\\

\begin{figure}[!htb]
\centering
\includegraphics[scale=0.5]{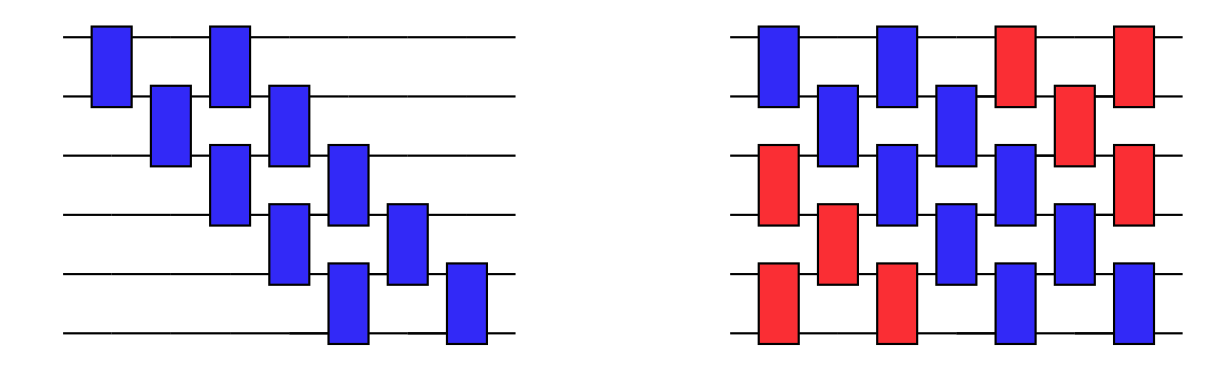}
\caption{Staircase circuit (left) vs brickwork circuit (right): the staircase circuit has two ``staircase layers". Its circuit depth increases linearly with the number of qubits. The brickwork circuit has the same depth as the staircase circuit but has more gates and thus more expressivity. The additional gates in the brickwork circuit  (shown in red) do not increase the circuit depth with respect to the staircase circuit.}\label{stair_vs_brick}
\end{figure}

\begin{figure}
\centering
\includegraphics[scale=0.7]{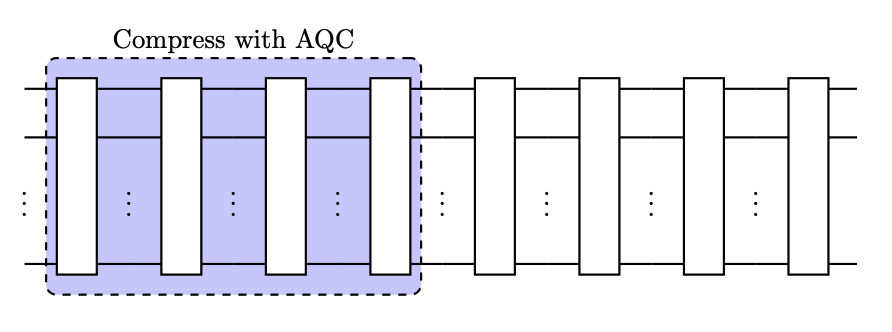}
  \caption{Schematic of our approach: Trotterisation is applied classically (purple box) using TEBD and then a Matrix Product State implementation of Approximate Quantum Compiling is applied to compress this part of the circuit. Standard Trotterisation is then applied on a quantum  device afterwards to simulate longer times, i.e. times that are beyond what is possible classically.}\label{schematic}
\end{figure}

\noindent All of these algorithms, including the one found to have the best performance, use a ``staircase circuit" with a linear structure - see Figure \ref{stair_vs_brick}. The depth of a ``staircase circuit" scales linearly with the number of qubits, thus the minimum depth of even the \textit{least} expressive $100$-qubit parametric circuit of this type would already be deeper than some of the deepest circuits that have been executed to date \cite{kim2023evidence}. By using a more expressible Ansatz such as the brickwork circuit in Figure \ref{stair_vs_brick}, we obtain a circuit with significantly more expressivity but with the same circuit depth. As we will show, using the brickwork circuit also allows to use a ``smart-initialisation" scheme in the optimization procedure. We note that the linear scaling of the depth with qubit number of the staircase circuit is a necessary requirement if one's aim is to prepare a quantum state whose lightcone increases with the number of qubits. However, if one considers a Matrix Product State with a bond dimension $\chi$ that is constant in the number of qubits, then it is not necessary to increase the circuit depth as one increases the number of qubits. An example of such a scenario is the preparation of a quantum state generated by the time evolution of a local Hamiltonian at a fixed time - the main focus of this work.
We will show that our $AQCtensor$ algorithm can be used to significantly reduce the depth of quantum circuits used to simulate the time evolution of quantum many-body systems, even beyond the regime of classical simulability. The workflow to achieve this is shown in Figure \ref{schematic} and discussed briefly below. \\

\noindent First, one uses a classical algorithm to store a time-evolved quantum state classically as a Matrix Product State. For a one-dimensional system with local interactions one can use e.g. TEBD \cite{hauschild2018efficient, vidal2004efficient} to prepare the time-evolved MPS, while for models in higher dimensions or with long-range interactions one can use other methods such as the TDVP algorithm \cite{haegeman2011time, haegeman2016unifying} or the $W^{I, II}$ method \cite{zaletel2015time}. The bond dimension $\chi$ typically increases exponentially with the time that is simulated, thus one eventually reaches a time $t$ at which the classical capabilities are saturated. Second, one uses $AQCtensor$ to find a short-depth circuit that prepares this state. Finally, one then applies additional time steps, e.g. using Trotterization, to prepare a state that, if stored as a Matrix Product State, would have a higher bond dimension than the classical computing resources could store. We will show that this procedure leads to a quantum circuit with a depth which is significantly lower than if one only used a Trotterization scheme without applying $AQCTensor$, or if one used one of the previously introduced MPS to quantum circuit mappings.\\

\noindent We note that a number of schemes to optimize parametric circuits on a quantum device to find circuits that simulate time evolution with depths lower than Trotterized circuits have been proposed \cite{gibbs2022long, cirstoiu2020variational, zoufal2023error, bharti2021quantum}. While these approaches offer a number of insights, particularly regarding error bounds \cite{zoufal2023error}, each of them suffers from a number of issues such as convergence, runtime and limited device connectivity. As a result, it has been argued that such quantum-variational approaches are not practical for use on near term quantum hardware \cite{miessen2021quantum}. Our approach, however, performs the optimization procedure \textit{classically} and hence avoids these issues, allowing us to run the algorithm on 100 qubits - see main text.\\

We briefly comment on $AQCTensor$ in the context of other compilation algorithms for the purposes of Hamiltonian simulation. $AQCTensor$ can be applied in tandem with these approaches as a procedure to prepare some non-trivial initial state. Consider for example the works \cite{tepaske2023, lubasch2023} which use Matrix Product Operators to optimize for the full unitary representing the time evolution operator. One can apply the workflow presented here to generate a short depth circuit that prepares the state at the largest time allowed by the available classical resources, followed by the application of the optimized circuits representing the time evolution operator. We point out that $AQCTensor$ finds an optimal circuit with respect to the initial state \textit{and} the time evolution operator - using both methods simultaneously would thus be advantageous. Although we note that the classical resources required to optimize for the full unitary are quadratically larger than those required to optimize for the quantum state. One can adopt the same approach for other simulation algorithms such as QDrift \cite{campbell2019random} and multiproduct formulas \cite{vazquez2023well, zhuk2024trotter}.\\

\noindent The structure of this work is as follows: in section \ref{background} we review the classical and quantum time evolution algorithms based on Trotterization as well as a brief introduction to Matrix Product States and how they are used in this work. In section \ref{sec:algorithm} we describe the $AQCtensor$ algorithm, leaving some details to the appendix. In section \ref{sec:results} we present the results of our approach applied to spin chains with $100$ qubits - see Figure \ref{fig:results} where the performance of the optimized circuits are compared with standard Trotterized circuits. We discuss the application of the method to models in higher dimensions and with longer range interactions in Appendix \ref{app:beyond_1D} - see Figures \ref{fig:long_range_fid} and \ref{fig:long_range_convergence}. We conclude in section \ref{sec:discussion} with a discussion of extensions of this work and how it can be combined with other methods.

\section{Background}\label{background}
\subsection{Time Evolution}
We first consider a family of spin chains with Hamiltonians of the form:
\begin{equation}\label{hxyz}
    H_{XYZ} = -\sum_{i=0}^{L-2}\left( \alpha_i S^x_i S^x_{i+1} + \beta_i S^y_i S^y_{i+1} + \Delta_i S^z_i S^z_{i+1}  \right) + \sum\limits_{i=0}^{L-1}h_iS^z_i,
\end{equation}
where $S^x$, $S^y$ and $S^z$ are written in terms of Pauli matrices as $S^x = \frac{\sigma^x}{2}$, $S^y = \frac{\sigma^y}{2}$ and $S^z = \frac{\sigma^z}{2}$. We will consider models with long range interactions in appendix \ref{app:beyond_1D}. Note that we can recover the XXZ and XXX models - prototypical examples of 1D integrable models - by setting $\alpha_i = \beta_i = 1$ or $\alpha_i = \beta_i = \Delta_i = 1$ respectively. The dynamical behaviour of these models has been studied extensively  \cite{piroli2017quantum}, including on a quantum computer \cite{smith2019simulating, keenan2023evidence}.\\

\noindent The time evolution of a quantum state $\ket{\psi(t)}$ is governed by the Schr{\"o}dinger equation:
\beq\label{schrod_eq}
 \ket{\psi(t)}= e^{-iH_{XYZ}t} \ket{\psi(0)}
\eeq
where $\ket{\psi(0)}$ is the wavefunction at time $t=0$. We will consider the initial state to be a product state such as e.g. the N{\'e}el state, written as: $\ket{\uparrow\downarrow\uparrow\downarrow...\uparrow\downarrow}$ where $\uparrow$ and $\downarrow$ represent up and down spins respectively. The N{\'e}el state for $n$ spins is simply implemented on $n$ qubits as $\ket{1010...10}$. \\

\begin{figure}
    \centering
    \includegraphics[scale=0.7]{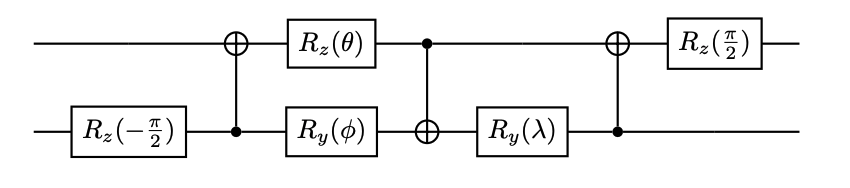}
    \caption{Implementation of two site operator $e^{i(\alpha S^x \otimes S^x + \beta S^y \otimes S^y + \Delta S^z \otimes S^z)}$ as a quantum circuit. We have the correspondences $\theta = \frac{\pi}{2}-\frac{1}{2}\Delta$, $\phi = \frac{1}{2}\alpha-\frac{\pi}{2}$ and $\lambda=\frac{\pi}{2}-\frac{1}{2}\beta$.}
    \label{fig:ut}
\end{figure}

\begin{figure}
    \centering
    \includegraphics[scale=0.5]{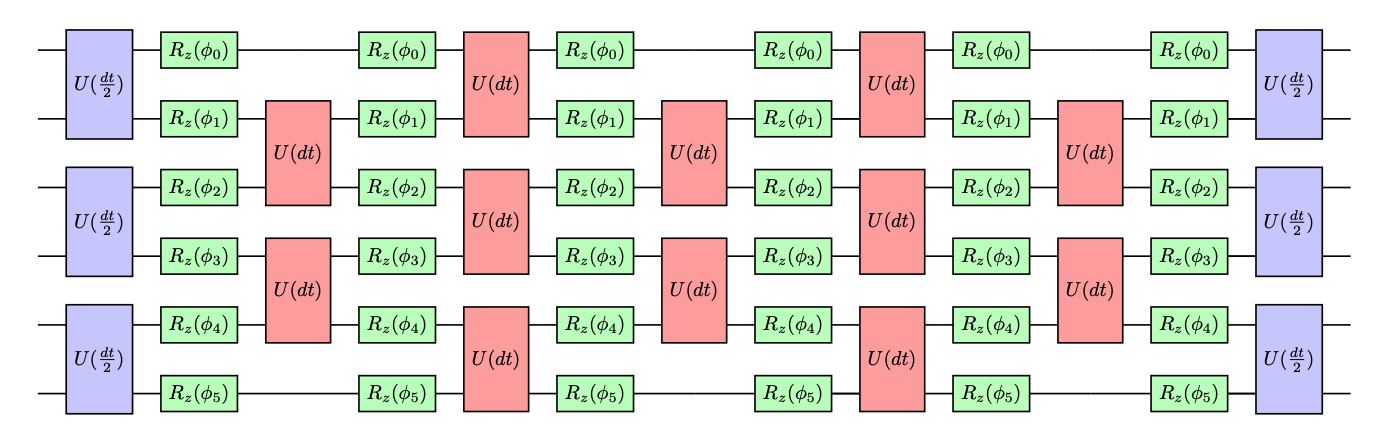}
    \caption{Second order Trotter circuit acting on six qubits. The single qubit rotation gates are shown in green and implement the terms of the form $e^{-\frac{1}{2}ih_jS^z dt}$ in equation (\ref{utrott2}). These rotation gates take angles given by $\phi_i = h_idt$. The purple two-qubit gates implement the half Trotter steps - i.e. terms of the form $U_{2j-1, 2j}(\frac{dt}{2})$ in equation (\ref{utrott2}). The red two-qubit gates implement full Trotter steps given by $U_{2j-1, 2j}(dt)$.}
    \label{fig:trott2}
\end{figure}

\noindent The time evolution operator $U(t)\equiv e^{-iH_{XYZ}t}$ can be executed as a quantum circuit in a resource efficient way; we first define $h_{i, i+1} \equiv \alpha_i S^x_i S^x_{i+1} + \beta_i S^y_i S^y_{i+1} + \Delta_i S^z_i S^z_{i+1} $ and then rewrite the Hamiltonian in (\ref{hxyz}) as $H_{XYZ} = H_1 + H_2 + \sum\limits_{i=0}^{L-1}h_iS^z_i$ where $H_1 = -\sum\limits_{i\text{ odd}} h_{i, i+1}$ and $H_2 = -\sum\limits_{i\text{ even}} h_{i, i+1}$. Next we define the unitary $U_{j, j+1}(dt) = e^{ih_{j, j+1}dt}$ for some time step $dt$. We can then define a symmetric, second-order Suzuki-Trotter time evolution operator $\mathcal{U}_{\text{trott}}(dt)$ in the following way:
\beq\label{utrott2}
\begin{aligned}
    \mathcal{U}_{\text{trott}}(dt) = \prod_{j=0}^{L/2-1}U_{2j, 2j+1}\left(\frac{dt}{2}\right) \prod_{j=0}^{L-1} e^{-\frac{1}{2}ih_jS^z dt} & \prod_{j=1}^{L/2-1}U_{2j-1, 2j}\left(dt\right) \prod_{j=0}^{L-1} e^{-\frac{1}{2}ih_jS^z dt} \prod_{j=0}^{L/2-1}U_{2j, 2j+1}\left(\frac{dt}{2}\right)\\
\end{aligned}
\eeq
The Suzuki-Trotter time evolution operator $\mathcal{U}_{\text{trott}}(dt)$ is an approximation of the exact time evolution operator $U(t)\equiv e^{-iH_{XYZ}t}$, in particular we have:
\beq\label{trott_error}
\mathcal{U}_{\text{trott}}(dt) = e^{-iH_{XYZ}t} + O(dt^3)
\eeq
As discussed in \cite{smith2019simulating}, each $U_{j, j+1}(dt)$ can be implemented by the quantum circuit with just three CNOTs as in Figure \ref{fig:ut}. The full second-order Suzuki-Trotter time evolution operator $\mathcal{U}_{\text{trott}}(dt)$ can be implemented as a quantum circuit in Figure \ref{fig:trott2}. Note that each ``half Trotter step" $U_{2j, 2j+1}\left(\frac{dt}{2}\right)$ in equation (\ref{utrott2}) can be fused into a full Trotter step from the previous layer, so the half steps only appear at the beginning and at the end of the circuit - see the purple gates in Figure \ref{fig:trott2}.

\subsection{Matrix Product States}\label{sec:mps}
\begin{figure}
\centering
\includegraphics[scale=0.7]{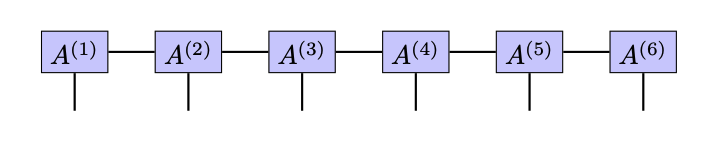}
\caption{Graphical representation of an MPS. There are two matrices $A^{(i)}$ for each qubit at position $i$.}\label{fig:mpschain}
\end{figure}

In this section we briefly describe the properties of Matrix Product States. For more details, we refer the reader to more in depth reviews \cite{ran2020tensor, schollwock2011density, evenbly2022practical}. An arbitrary quantum state on $n$ qubits can be written in terms of complex variables $c_{j_1,...,j_n}$, the number of which scales as $2^n$:
\beq
\ket{\psi} = \sum\limits_{\{j_1,...,j_n \}} c_{j_1,...,j_n}\ket{j_1,...,j_n}
\eeq
where the sum is over all configurations of the binary variables $j_1,...,j_n$. The bipartite entanglement entropy of an arbitrary quantum state picked at random from Hilbert space satisfies a volume law which is distinct from area law entanglement in which case the entanglement entropy of two regions after the bipartition of the system is proportional to the area of the boundary of the system. A small subset of states in Hilbert space satisfies an area law. The coefficients $c_{j_1,...,j_n}$ of such states have a certain structure that we can take advantage of to study classically. Any state $\ket{\psi}$ can be written in the following way:
\beq\label{eq:mps}
c_{j_1,...,j_n} = A_{j_1}^{(1)}\cdot A_{j_2}^{(2)}...\cdot A_{j_n}^{(n)}
\eeq
where the $A_j$ are $\chi_j \times \chi_{j+1}$ dimensional matrices. Quantum states of the form (\ref{eq:mps}) are known as Matrix Product States (MPS). The maximum value of $\chi_j$ is referred to as the bond dimension of the MPS. We can represent an MPS graphically as in Figure \ref{fig:mpschain}. The bond dimension $\chi_j$ can be seen as a measure of the entanglement between the two subsystems when a bipartition is made at qubit $j$. Therefore, states in Hilbert space that satisfy an area law - and therefore have a low bond dimension in their MPS representation - can be efficiently stored as Matrix Product States. States that satisfy a volume law will have a bond dimension that is exponential in the number of qubits. We will consider in this work the non-trivial dynamics governed by equation (\ref{schrod_eq}). The bipartite entanglement entropy of a ground state of a one-dimensional Hamiltonian that has a gap between its ground state and its excited state is independent of the size of the subsystems \cite{hastings2007area}. The ground state of such a system - and hence the initial state in our setup - can be efficiently stored as an MPS. One can then use an algorithm such as TEBD (Time Evolving Block Decimation) \cite{hauschild2018efficient, vidal2004efficient} or TDVP \cite{haegeman2011time, haegeman2016unifying} to update the MPS as a function of time to study the dynamics of the system. However, the entanglement entropy of the state increases linearly with time, hence the bond dimension $\chi$ that is required to keep the error constant diverges exponentially with time, thus motivating the use of a quantum computer.

\begin{figure}
\centering
\includegraphics[scale=0.7]{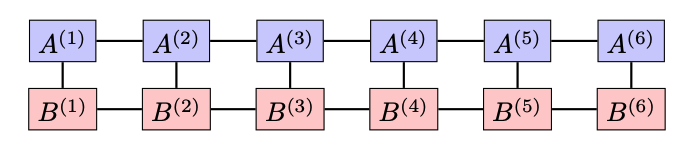}
\caption{The inner product $\braket{\psi_1| \psi_2}$ of two Matrix Product States - see equations (\ref{psi1psi2}) and (\ref{eq:mps}).}\label{fig:mpsnorm}
\end{figure}

\section{Algorithm}\label{sec:algorithm}
\begin{figure}
    \centering
    \includegraphics[scale=0.5]{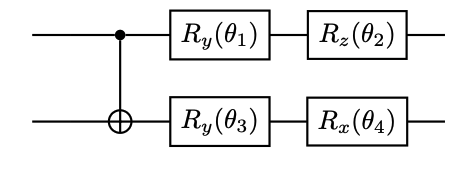}
    \caption{$CNOT$ block - each block corresponds to one of the purple two-qubit gates in Figure \ref{paramcircuit2}.}
    \label{fig:cnot_block}
\end{figure}
\begin{figure}
\centering
\includegraphics[scale=0.6]{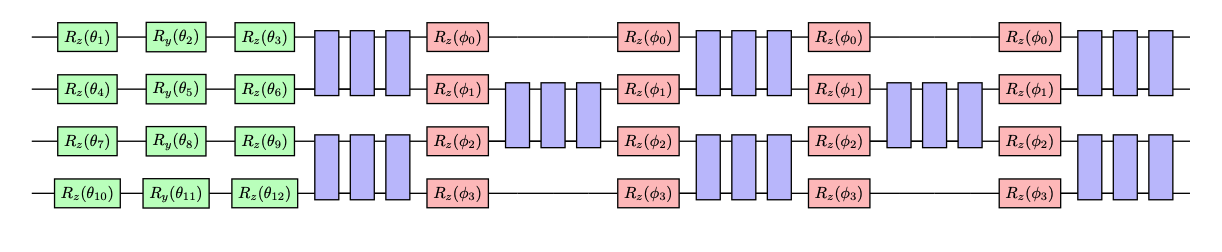}
  \caption{Parameterised circuit inspired by the structure of the second order Suzuki-Trotter circuit in Figure \ref{fig:trott2}. The angles of the red single qubit rotation gates are given by $\phi_i = h_idt$. The angles of the green single qubit gates are parameters to be optimized. The purple two-qubit gates are $CNOT$ blocks shown in Figure \ref{fig:cnot_block}, each with an independent set of parameters $\theta_1, \theta_2, \theta_3, \theta_4$ so that there is a total of $108$ free parameters in this circuit - $12$ in the green gates and $96$ in the purple gates. Note that in each triplet of purple $CNOT$ blocks, the direction of the $CNOT$ gate in the outer two blocks should be reversed in order to match exactly the $CNOT$ structure of the Suzuki-Trotter circuit in Figure \ref{fig:trott2}.
  }\label{paramcircuit2}
\end{figure}

\begin{algorithm}
    \caption{$AQCtensor$ algorithm}\label{alg:alg_details}
    \hspace*{\algorithmicindent} \textbf{Input} Hamiltonian $H$, initial state $\ket{\psi(0)}$, parametric circuit $V(\theta_0)$ \\
    \hspace*{\algorithmicindent} \textbf{Output} Optimized parametric circuit with appended Trotter steps: $(U_{trott}(dt))^k V(\theta)$ 
    \begin{algorithmic}
    \State 1. Apply classical algorithm (e.g. TEBD) to generate state $\ket{\psi(t)}$
    \State 2. Minimize cost function in (\ref{cost_func_alpha2}) with respect to parameters $\theta$ to find optimized parameters $\theta_0$
    \State 3. Append $k$ additional Trotter steps with time step $dt$ to $V(\theta_0)$ to obtain the circuit $(U_{trott}(dt))^k V(\theta_0)$
    \end{algorithmic}
\end{algorithm}
Approximate quantum compiling (AQC) involves the design of a parametric quantum circuit with fixed depth - the parameters are then adjusted to bring it as close as possible to the target, where “close” is defined via some carefully chosen metric, see below. A parametric circuit is constructed from $CNOT$ blocks, each of which has four parameters - see Figure \ref{fig:cnot_block}.
The position of the $CNOT$ blocks in the parameterised circuit can be customised to suit the particular target circuit that one is interested in. Here we are interested in finding a circuit that implements the unitary time evolution operator, which is written in equation (\ref{schrod_eq}) for the case of a one-dimensional spin chain. For this family of models, we thus consider a structure inspired by the second-order Trotter circuit in Figure \ref{fig:trott2}. Recall that each block $U(dt)$ in Figure \ref{fig:trott2} represents the 2-qubit sub-circuit with three $CNOT$s in Figure \ref{fig:ut}; it is therefore natural to consider a circuit Ansatz with sub-circuits each with three $CNOT$ blocks as in Figure \ref{paramcircuit2}, such that the circuit Ansatz mimics the structure of the second order Trotter circuits. We denote the number of repetitions of $CNOT$ blocks in each layer by $3$, and the number of layers by $l$ - the parameterised circuit in Figure \ref{paramcircuit2} corresponds to $n=4$ qubits, $l=2$ layers and $b=3$ $CNOT$ blocks in each layer. Note that when we take $b=3$, the total $CNOT$ depth of the parameterised circuit with $l$ layers is equal to the $CNOT$ depth of a Trotter circuit with $l$ Trotter steps - see Figure \ref{fig:trott2}. In Figure \ref{paramcircuit2} there are three rotation gates acting on each qubit at the beginning of the circuit. If the initial state of the parametric circuit is $\ket{0}$ then the initial rotation gate $R_z(\theta)$ is redundant but would act non-trivially on more general initial states. \\

\noindent One can define the distance between the target and parameterised circuit via a number of different metrics. Here we define this distance based on the Hilbert-Schmidt test:
\beq\label{cost_state_prep}
C_{hs}^{\text{state}} = 1 -  |\bra{0}V^{\dagger}(\theta)\ket{\psi_t}|^2 
\eeq
where $V(\theta)$ is the unitary operator corresponding to the parametric circuit. Note that here we are considering the application of AQC to state preparation as opposed to full circuit compilation. More precisely, this means that our cost function is designed such that it is minimised when the action of $V(\theta)$ on the initial state produces a state that is as close as possible to a target state $\ket{\psi_t}$ (up to some global phase). This is in contrast to the situation where one starts with some target \textit{circuit} $U$ and the cost function is designed to bring the full matrix $V(\theta)$ as close as possible to $U$. \\

It was pointed out in \cite{khatri2019quantum} that the gradient of the cost function in (\ref{cost_state_prep}) vanishes exponentially in the number of qubits. While this is primarily a problem when training on a quantum device - which is not the case in the $AQCtensor$ framework - we consider problem sizes of up to to 100 qubits, thus the exponentially vanishing gradient of  $C_{hs}^{\text{state}}$ can lead to impractically long training times even when using classical methods to train the circuit parameters. This is related to the so-called ``orthogonality catastrophe" whereby the overlap of two quantum states in Hilbert space vanishes exponentially in the number of qubits, resulting in $C_{hs}^{\text{state}} \approx 1$ for a large region of parameter space. Here we will briefly introduce two methods to overcome this issue - more details can be found in appendix \ref{app:circ_opt}. The first method is based on the concept of ``local cost functions" \cite{ khatri2019quantum, robertson2022escaping} and involves the addition of terms to $C_{hs}^{\text{state}}$, so that our algorithm uses a cost function of the form:
\beq\label{cost_func_alpha2}
C_L^{(1)}(\alpha) = 1 - |\bra{0}V^{\dagger}(\theta)\ket{\psi_t}|^2 -
\alpha\sum\limits_{j=1}^n| \bra{0}X_j V^{\dagger}(\theta)\ket{\psi_t}|^2
\eeq
where $\alpha$ is a parameter that can be tuned throughout the optimization procedure. We discuss in appendix \ref{app:circ_opt} how the additional term on the RHS of (\ref{cost_func_alpha2}) arises from expanding and truncating the full expression for the local cost function introduced in \cite{khatri2019quantum}. The second method that we introduce is ``Trotter-initialization" - whereby we tune the initial values of the circuit parameters, i.e. at the beginning of the optimization procedure, to take values such that the parametric circuit initially corresponds exactly to the Trotter circuit. This has the advantage that performing even one step of the optimization procedure ensures that our parametric circuit has a higher fidelity with the exactly time-evolved state than a standard Trotter circuit. Note that the ``Trotter-initialization" method is only possible when the $CNOT$-structure of the parametric circuit matches that of the Trotter circuit, as is the case for the circuit in Figure \ref{paramcircuit2}.\\

\noindent Note that each term in (\ref{cost_func_alpha2}) is an overlap of quantum states, and since the overlap of two MPS can be calculated very efficiently, the architecture of Matrix Product States can be leveraged to calculate the cost function and solve the approximate quantum compilation problem for large numbers of qubits. Consider for example two quantum states $\ket{\psi_1}$ and $\ket{\psi_2}$:
\beq\label{psi1psi2}
\begin{aligned}
\ket{\psi_1} &= \sum\limits_{\{j_1,...,j_n \}} c^{(1)}_{j_1,...,j_n}\ket{j_1,...,j_n}\\
\ket{\psi_2} &= \sum\limits_{\{j_1,...,j_n \}} c^{(2)}_{j_1,...,j_n}\ket{j_1,...,j_n}\\
\end{aligned}
\eeq
As discussed in section \ref{sec:mps}, for weakly entangled states the coefficients $c^{(1)}_{j_1,...,j_n}$ and $c^{(2)}_{j_1,...,j_n}$ can be represented efficiently as Matrix Product States:
\beq\label{eq:mps2}
\begin{aligned}
c_{j_1,...,j_n}^{(1)} &= A_{j_1}^{(1)}\cdot A_{j_2}^{(2)}...\cdot A_{j_n}^{(n)}\\
c_{j_1,...,j_n}^{(2)} &= B_{j_1}^{(1)}\cdot B_{j_2}^{(2)}...\cdot B_{j_n}^{(n)}
\end{aligned}
\eeq
The fidelity of two MPS:
\beq\label{eq:mpsfid}
f(\ket{\psi_1}, \ket{\psi_2}) = ||\braket{\psi_1 | \psi_2}||^2
\eeq
can be calculated efficiently by contracting the Tensor Network - see Figure \ref{fig:mpsnorm}. One has a choice regarding the definitions of the states $\ket{\psi_1}$ and $\ket{\psi_2}$ when evaluating the cost function in (\ref{cost_func_alpha2}). One can for example define $\ket{\psi_1} = V(\theta)\ket{0}$ and $\ket{\psi_2} = \ket{\psi_t}$ or $\ket{\psi_1} = \ket{0}$ and $\ket{\psi_2} = V^{\dagger}(\theta)\ket{\psi_t}$. The latter approach has the advantage that, when the parameters $\theta$ are close to their optimal values then both $\ket{\psi_1}$ and $\ket{\psi_2}$ are weakly entangled states, thus allowing us to reduce the errors induced from the truncation of the bond dimension. We thus adopt this approach in our simulations, but we point out that for some other families of Hamiltonians this may not necessarily be the optimal strategy.

\section{Results}\label{sec:results}
\begin{figure}[!htb]\centering
\centering
\includegraphics[scale=0.3]{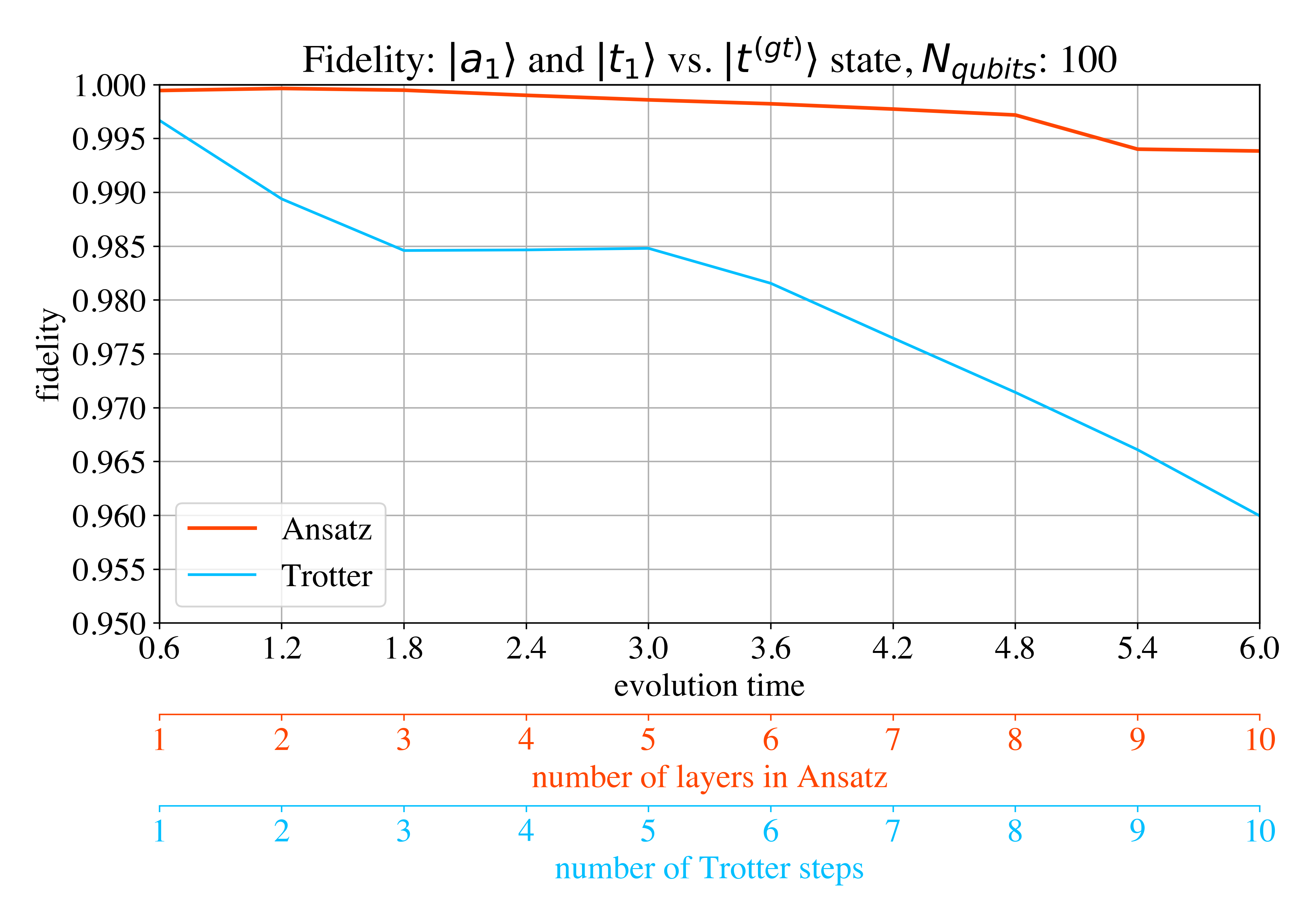}
\includegraphics[scale=0.3]{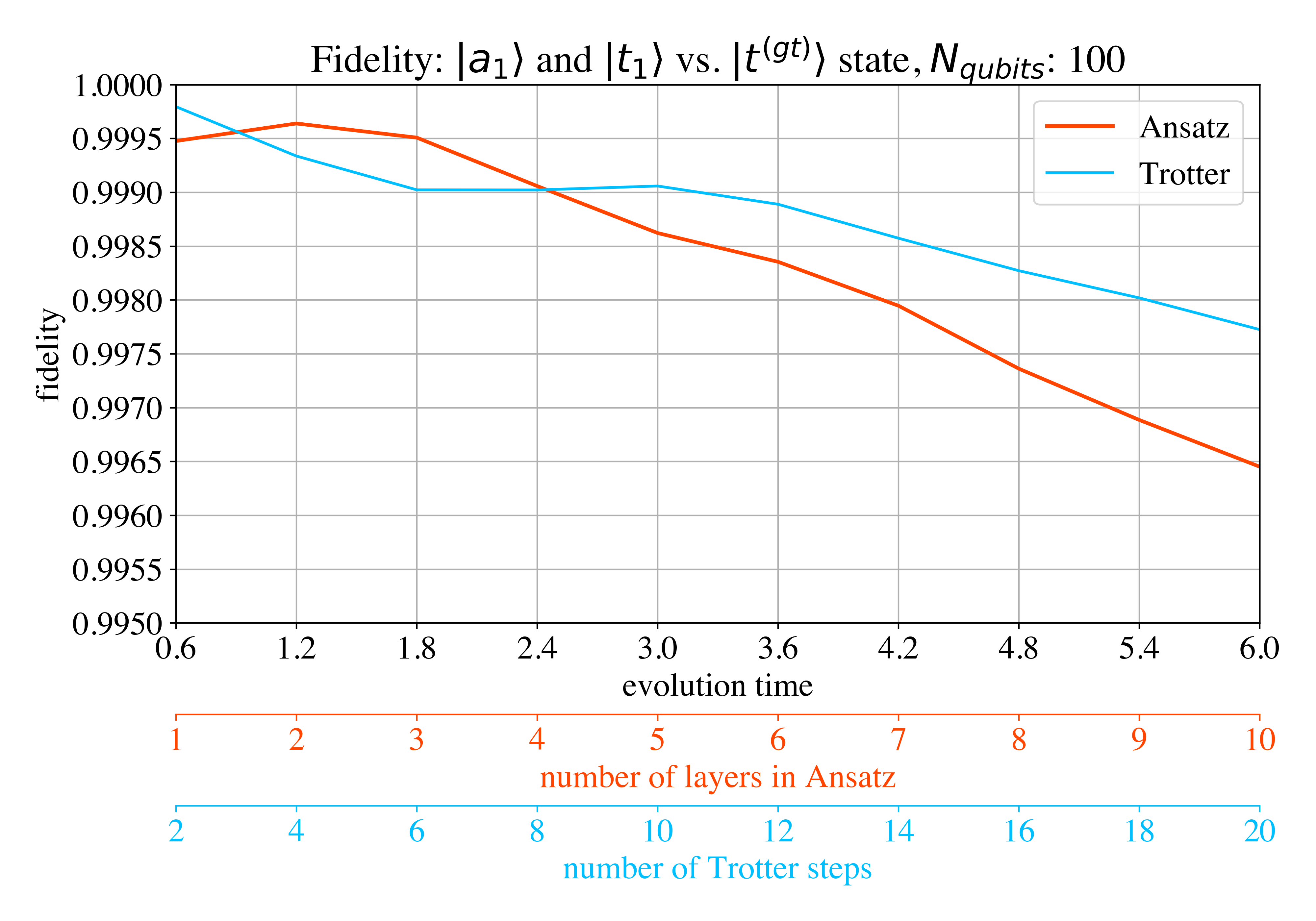}
\includegraphics[scale=0.3]{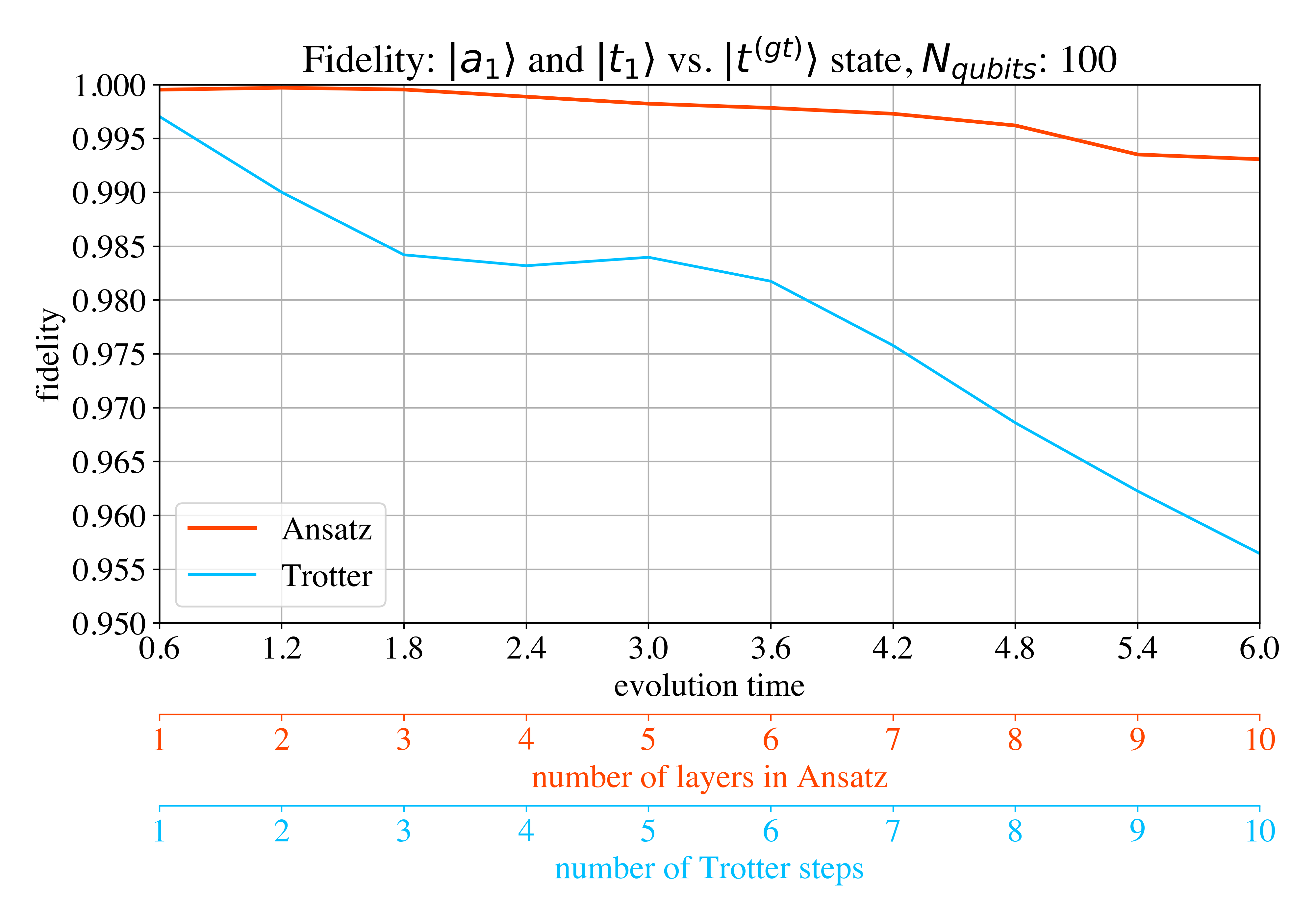}
\includegraphics[scale=0.3]{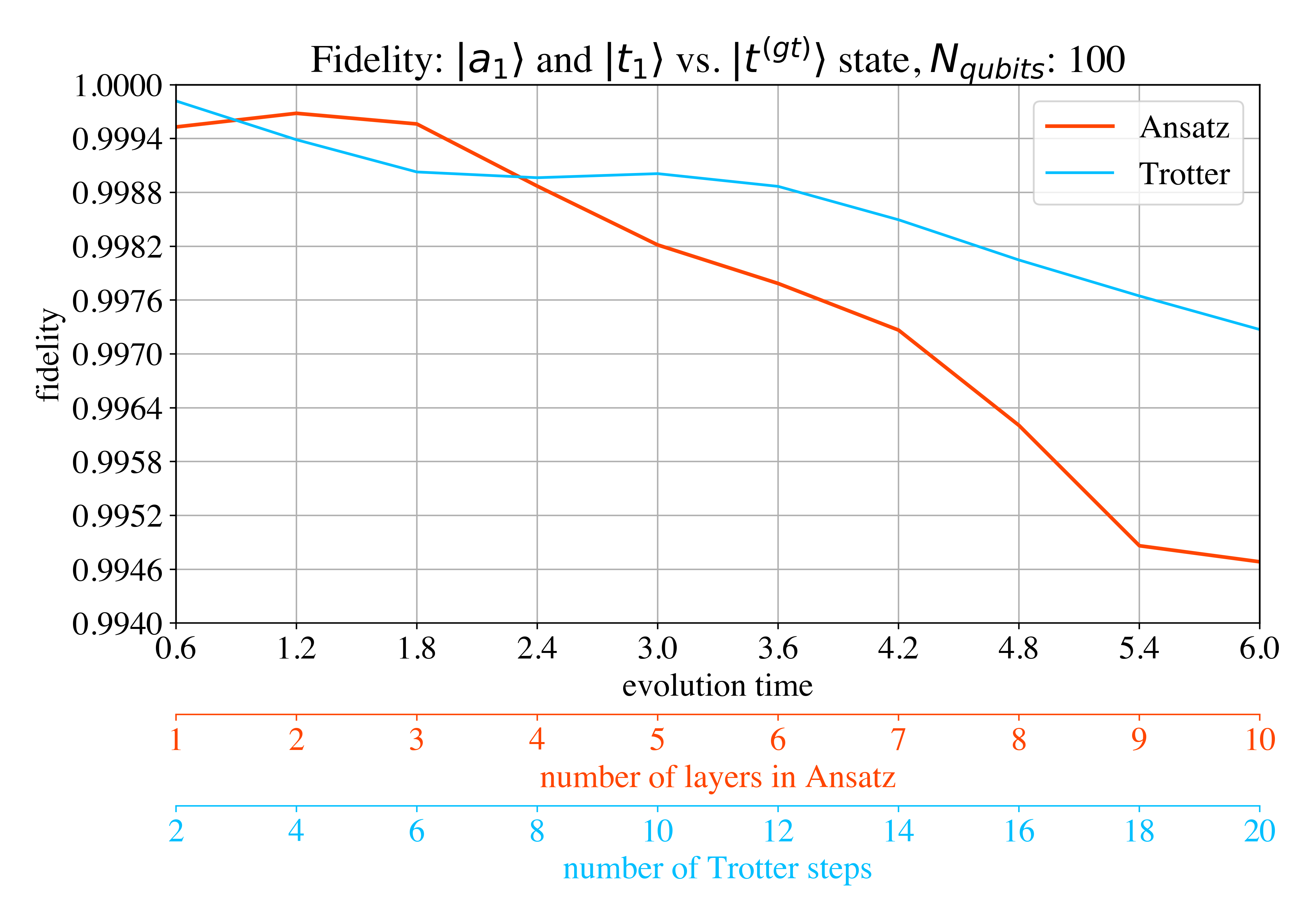}
\includegraphics[scale=0.3]{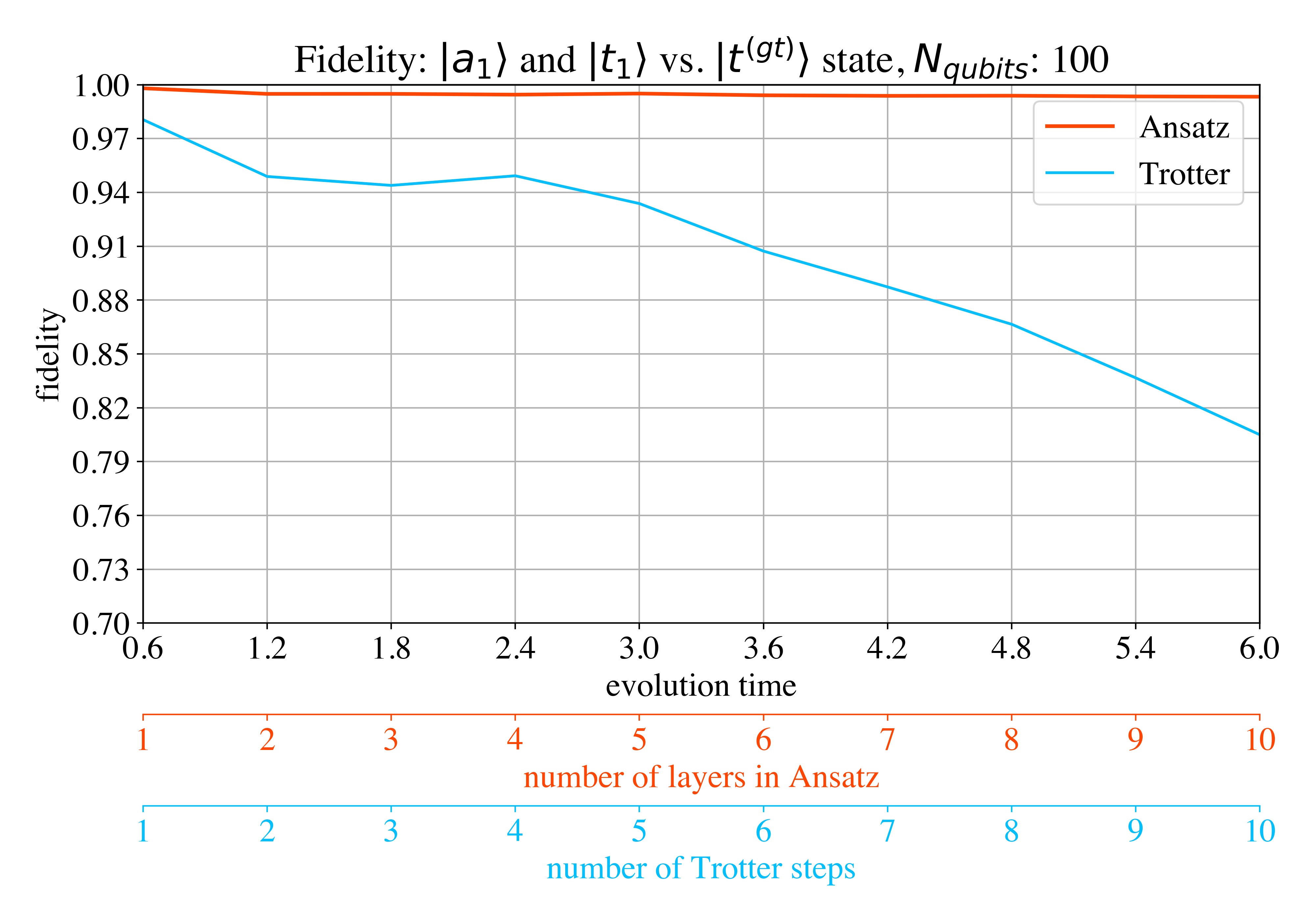}
\includegraphics[scale=0.3]{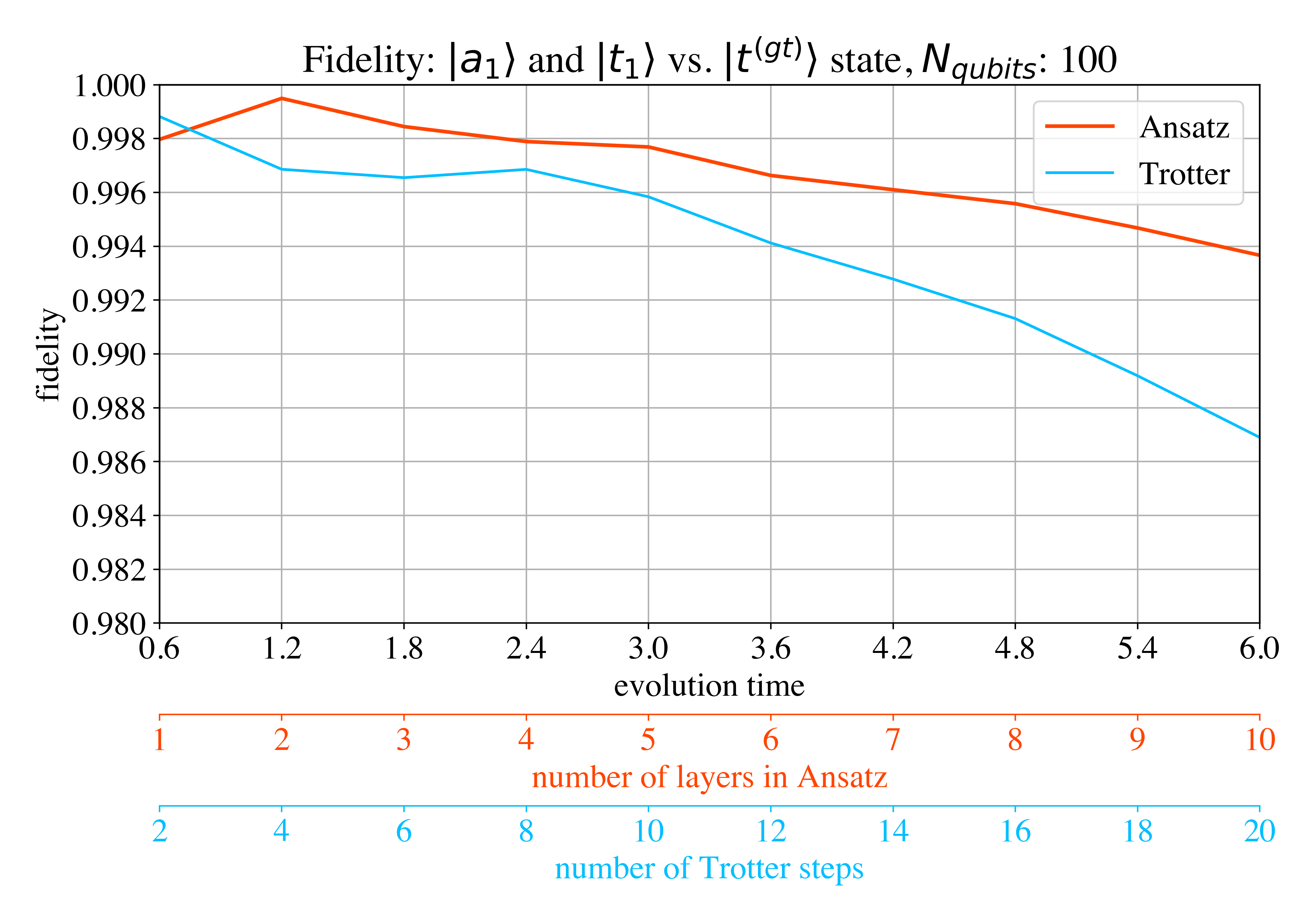}
\caption{Top panel: $H = -\sum_{i=0}^{L-1}\left( \alpha_i S^x_i S^x_{i+1} + \beta_i S^y_i S^y_{i+1} + \Delta_i\, S^z_i S^z_{i+1} \right)$ with $\alpha_i, \beta_i, \Delta_i$ taking randomly generated values between $0.375$ and $1.125$. Middle panel: $H = -\sum_{i=0}^{L-1}\left( 0.75\, S^x_i S^x_{i+1} + 0.75\, S^y_i S^y_{i+1} + 0.75\, S^z_i S^z_{i+1} \right)$. Lower panel: $H = -\sum_{i=0}^{L-1}\left( 0.75\, S^x_i S^x_{i+1} + 0.75\, S^y_i S^y_{i+1} + 1.5\, S^z_i S^z_{i+1} \right)$. In the left panels, the accuracy of the optimized parametric circuit Ansatz is compared with the accuracy of the Suzuki-Trotter circuits with the same circuit depth. On the right, the Suzuki-Trotter circuits have twice the circuit depth as the optimized parametric circuit yet have comparable accuracy. The maximum number of optimization iterations was set to 30.}
\label{fig:results}
\end{figure}
We now present our results of the simulations using the $AQCtensor$ algorithm. In this section we consider the application of $AQCtensor$ to the family of Hamiltonians defined in equation (\ref{hxyz}) and we leave the discussion of longer range Hamiltonians to appendix \ref{app:beyond_1D}. We consider two paradigms; firstly we compare Trotterized circuits and optimized circuits of equal depth and measure their respective fidelities with the ``ground-truth" circuit, i.e. the Matrix Product State corresponding to the state generated by a Trotterized unitary with a very small time step $dt$. We will show that the fidelities of the optimized parametric circuits are significantly and consistently larger than the Trotterized circuits with equal circuit depth (left panel of Figure \ref{fig:results}). Secondly, we compare Trotterized circuits and optimized circuits with similar fidelities and compare their circuit depths (right panel of Figure \ref{fig:results}). We find that the optimized circuits can achieve roughly equal fidelities to the Trotterized circuits with half the circuit depth. We consider three Hamiltonians in the class of models defined in equation (\ref{hxyz}): 1) randomly generated coefficients 2) the XXX Hamiltonian and 3) the XXZ Hamiltonian - see the caption in Figure \ref{fig:results}. We define the following notation:
\begin{itemize}
    \item $\ket{a_1}$: The state generated by the optimized parametric circuit in Figure \ref{paramcircuit2}.
    \item Number of layers $l$ in Ansatz: in Figure \ref{paramcircuit2} there are $l=2$ layers.
    \item $\ket{t_1}$: The state generated by the Trotter circuit in Figure \ref{fig:trott2}. The time step used to generate this state is given by the total time divided by the number of layers/Trotter steps.
    \item Number of Trotter steps: the analogue of the number of layers in the Ansatz circuits. In Figure \ref{fig:trott2} there are $3$ Trotter steps.
    \item $\ket{t^{(gt)}}$: the ``ground truth" generated by classical Tensor Network simulations of deep Trotter circuits, i.e. small time steps. We take a time step $dt$ that is $10$ times smaller than the time step used to generate $\ket{t_1}$.
\end{itemize}
The simulations were run on Xeon E5-2699, v4, 2.2 GHz machine with 44/88 cores/threads and took several hours each. We used the Qiskit Aer MPS simulator - we believe that runtime can be decreased significantly by improving the performance of this simulator. We used the L-BFGS classical optimizer to minimize the cost function in our simulations. We found that convergence of the cost function was enhanced by prepending the L-BFGS optimizer with a few steps of the ADAM optimizer.

\section{Discussion}\label{sec:discussion}
This work demonstrates the power of combining state-of-the art classical simulation methods with quantum computing. Rather than viewing quantum computing and tensor networks as competing approaches, our $AQCtensor$ algorithm allows one to combine the power of both of these tools. While this work focused on Matrix Product States and a family of one-dimensional spin chains, future work could apply a similar workflow to higher dimensional models, where more sophisticated tensor network algorithms are used to optimize quantum circuits. Our work also demonstrates the effect of using a tailored approach to MPS-quantum circuit mappings. By using such an approach we found a much shallower circuit that generated the time-evolved MPS than if we had used one of the generic algorithms discussed in \cite{rudolph2023decomposition}. We focused here on optimizing circuits for time evolution, but one could imagine applying a similar tailored approach to the MPS/quantum approaches to ground-state search and quantum machine learning described in \cite{rudolph2023synergistic, huggins2019towards}. The circuit depth reductions achieved by $AQCTensor$ allow for the reduction of hardware noise due to the reduced number of quantum gates in the optimized circuit. We note that, in addition to the error reduction achieved by reducing the circuit depth, one can \textit{also} apply error mitigation methods \cite{van2023probabilistic, kim2023evidence} to the optimized circuit to further reduce hardware noise. We also point out that the $AQCTensor$ workflow is still likely to be applicable in the fault tolerant era since many algorithms such as Quantum Phase Estimation require a good initial state preparation.\\

As discussed in section \ref{sec:algorithm}, there is no unique definition of the pair of states $\ket{\psi_1}$ and $\ket{\psi_2}$ that are contracted to obtain the cost function in (\ref{cost_func_alpha2}). Each choice of a particular pair of states corresponds to a particular order of bond dimension truncations. We used one particular choice for our extensive study of one-dimensional spin chains, namely: $\ket{\psi_1} = \ket{0}$ and $\ket{\psi_2} = V^{\dagger}(\theta)\ket{\psi_t}$. However, for other Hamiltonians with long-range interactions or in higher dimensions, one may find that a different approach works better. We lay the groundwork for the exploration of the application of $AQCTensor$ to these more complex models in appendix \ref{app:beyond_1D} where we present some initial results on a model defined on a 2D decorated hexagonal lattice as well as a spin chain with next to nearest neighbour interactions. We demonstrate that the $AQCTensor$ workflow can still be applied successfully to these models. Future work can consider how to optimize the method for these more challenging cases. In addition to improving these MPS-based methods, one could also consider using higher dimensional tensor networks, i.e. beyond MPS, within the $AQCTensor$ workflow to further improve its performance when applied to higher dimensional Hamiltonians.\\

Finally, we briefly comment on how one could combine our approach to optimizing circuits for time-evolution with other approaches. The workflow in Figure \ref{schematic} allows us to find short depth circuits with low Trotter error that simulate the time-evolved state up until the latest time that the classical resources available can handle. To mitigate the Trotter errors arising from the remaining part of the circuit which has not been optimized, one could use other techniques such as multiproduct formulas \cite{zhuk2024trotter, vazquez2023well} - this will be the focus of future work. Additionally, one could consider using additional more resource-intensive tensor-network techniques to optimize the circuits at later times using the unitary compilation techniques described in \cite{lubasch2023, tepaske2023}.

\clearpage
\appendix

\section{$AQCTensor$ beyond 1D models}\label{app:beyond_1D}

\begin{figure}[!htb]\centering
\centering
\includegraphics[scale=0.6]{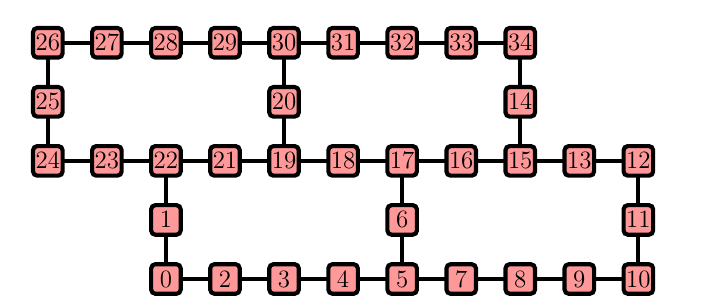}
\caption{The $2x2$ decorated hexagon graph on which the Hamiltonian in (\ref{h_hh}) is defined. To simulate this model with Matrix Product States, one first maps the two-dimensional topology to the chain by labelling each of the $35$ qubits with an integer. This results in the introduction of long-range interactions.}
\label{fig:hh_ordering}
\end{figure}

\begin{figure}[!htb]\centering
\centering
\includegraphics[scale=0.3]{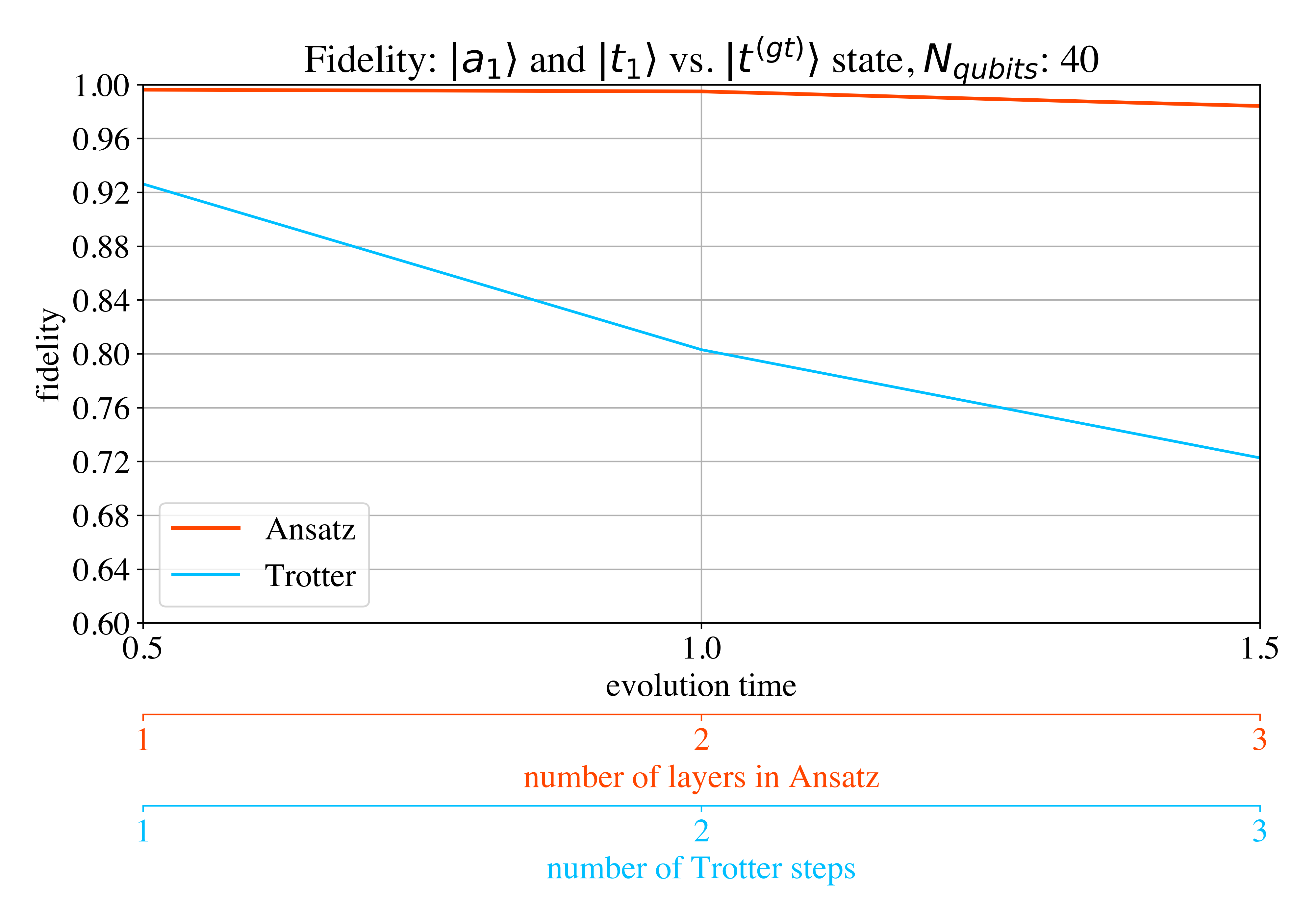}
\includegraphics[scale=0.3]{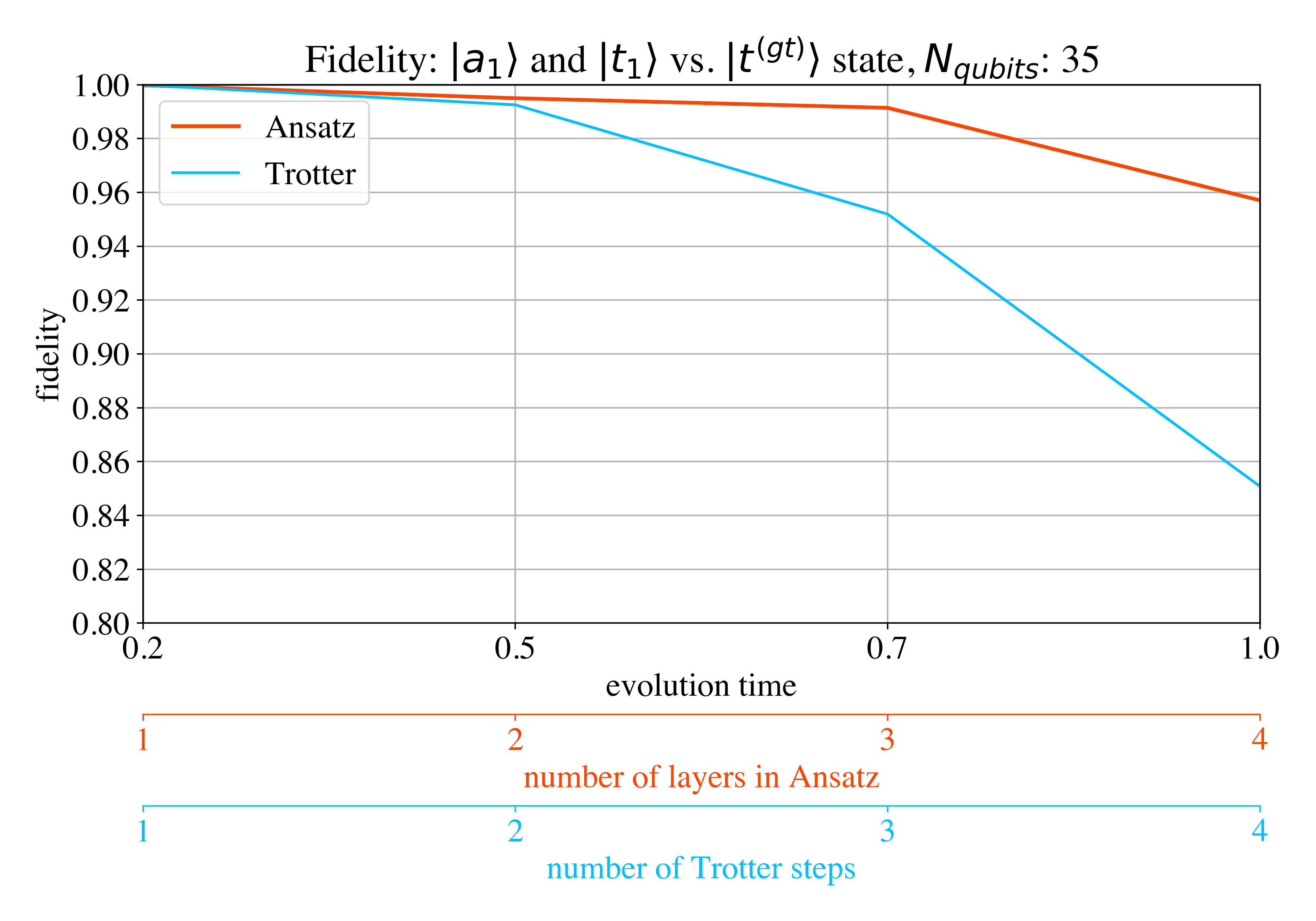}
\caption{Fidelities between the ground-truth circuits $\ket{t^{(gt)}}$ and the circuits optimized with $AQCTensor$. Left panel: the next to nearest neighbour Hamiltonian in (\ref{h_nnn}) with $\alpha_i=\beta_i=\Delta_i = 0.75$. Right panel: The Hamiltonian in (\ref{h_hh}) with $\alpha_i=\beta_i= 0.5$ and $\Delta_i=1.0$ with nearest neighbour couplings on the decorated hexagon lattice in Figure \ref{fig:hh_ordering}. }
\label{fig:long_range_fid}
\end{figure}

\begin{figure}[!htb]\centering
\centering
\includegraphics[scale=0.3]{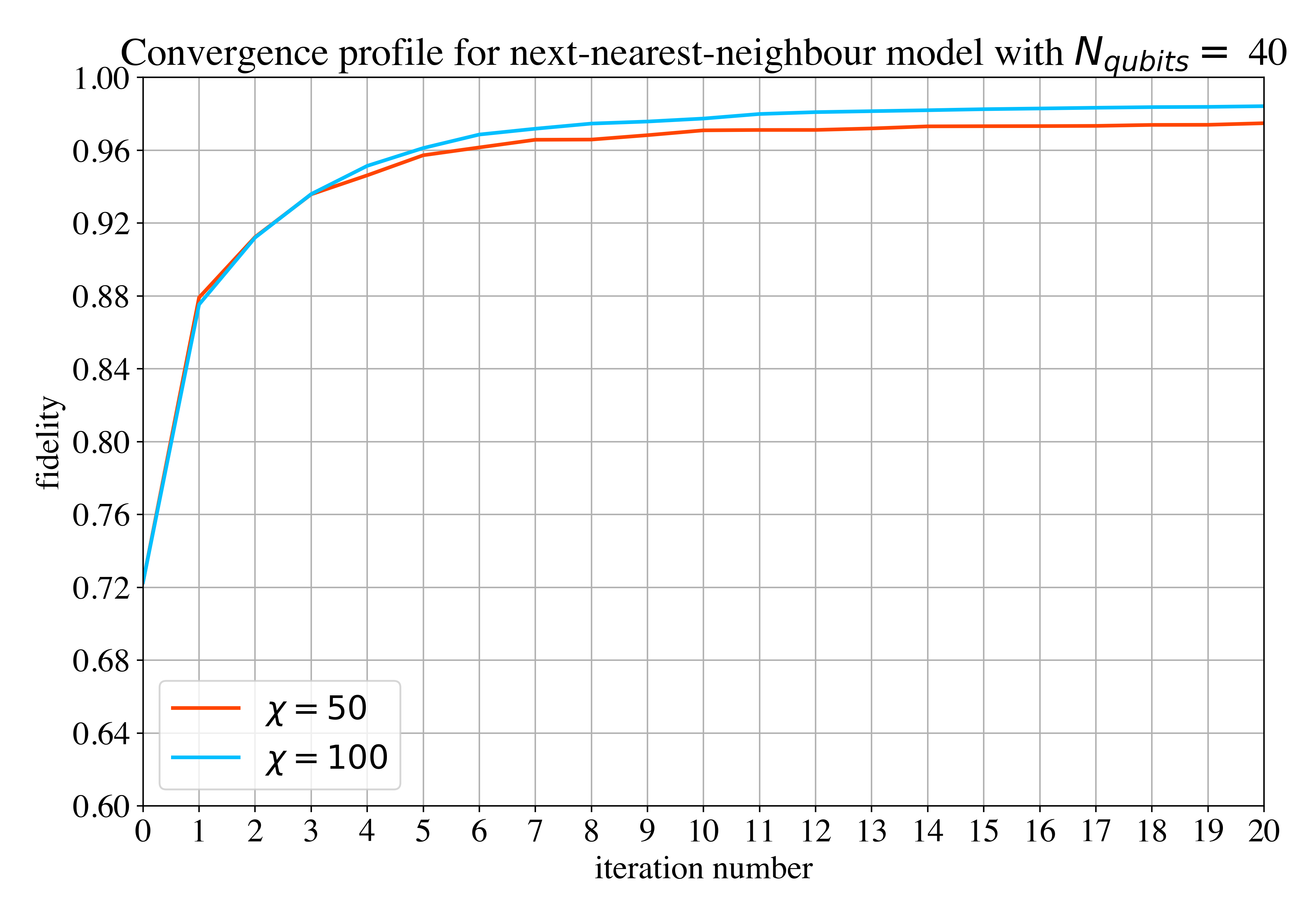}
\includegraphics[scale=0.3]{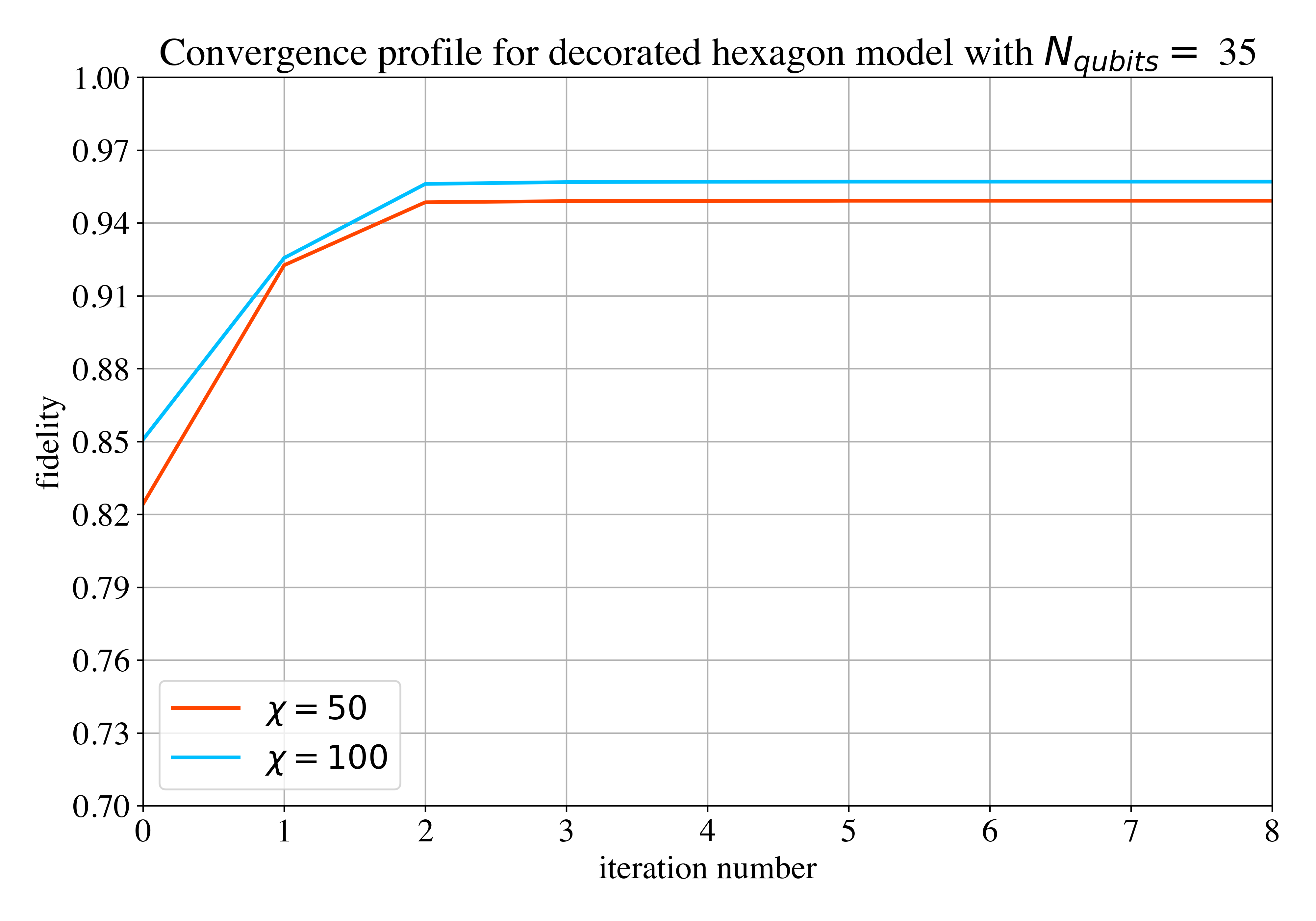}
\caption{The fidelities of the optimized circuits as a function of the iteration number in the optimization procedure. Left panel: the next to nearest neighbour Hamiltonian in (\ref{h_nnn}) at time $t=1.5$. Right panel: the decorated hexagonal Hamiltonian in (\ref{h_hh}) at time $t=1.0$.}
\label{fig:long_range_convergence}
\end{figure}

Here we consider two additional models to the family of Hamiltonians defined in equation (\ref{hxyz}). First, we consider a next to nearest neighbour Hamiltonian with Heisenberg interactions:
\begin{equation}\label{h_nnn}
    H_{nnn} = -\sum_{i=0}^{L-2}\left( \alpha_i S^x_i S^x_{i+1} + \beta_i S^y_i S^y_{i+1} + \Delta_i S^z_i S^z_{i+1}  \right) -\sum_{i=0}^{L-3}\left( \alpha_i S^x_i S^x_{i+2} + \beta_i S^y_i S^y_{i+2} + \Delta_i S^z_i S^z_{i+2}  \right)
    + \sum\limits_{i=0}^{L-1}h_iS^z_i,
\end{equation}
We take $\alpha_i=\beta_i=\Delta_i = 0.75$ for the purposes of our numerical studies below. Second, we consider a model defined on the decorated hexagonal lattice with Hamiltonian:
\begin{equation}\label{h_hh}
    H_{hex} = -\sum_{i,j \in e}^{L-2}\left( \alpha_i S^x_i S^x_{j} + \beta_i S^y_i S^y_{j} + \Delta_i S^z_i S^z_{j}  \right) 
    + \sum\limits_{i=0}^{L-1}h_iS^z_i,
\end{equation}
where the sum is over the edges $e$ in the graph such that $i,j$ are nearest neighbours on the decorated hexagonal lattice in Figure \ref{fig:hh_ordering}. For this model, we take $\alpha_i=\beta_i= 0.5$ and $\Delta_i=1.0$. We consider an array of $2\times2$ decorated hexagons which has a total of $35$ qubits. We apply MPS-based methods to this two-dimensional model by ordering the qubits as shown on the figure and `unrolling' the graph onto a one-dimensional spin chain. This has the effect of introducing long-range interactions - for example, qubits labelled $20$ and $30$ are nearest neighbours in Figure \ref{fig:hh_ordering} but will be separated by a distance of $10$ on the one-dimensional chain to which they are mapped. In Figure \ref{fig:long_range_fid}, we compare the fidelities with the ground truth circuit $\ket{t^{(gt)}}$ of the standard Trotterized circuits and the optimized circuits. It is observed that $AQCTensor$ significantly improves the fidelities for both the next to nearest neighbour model on $40$ qubits (left panel) and the $2\times2$ decorated hexagonal model on $35$ qubits (right panel). The cost function in both cases is evalauted using the choice $\ket{\psi_1} = \ket{0}$ and $\ket{\psi_2} = V^{\dagger}(\theta)\ket{\psi_t}$, as discussed in the main text. In Figure \ref{fig:long_range_convergence}, we plot the convergence of the fidelity as a function of the number of iterations in the optimization for the Hamiltonians in both (\ref{h_nnn}) and (\ref{h_hh}). We compare this convergence profile for two different values of the bond dimension in the MPS, $\chi=50$ and $\chi=100$. The results demonstrate that the optimization results have converged even at these very modest values of $\chi$. This is unsurprising as the states $\ket{\psi_1} = \ket{0}$ and $\ket{\psi_2} = V^{\dagger}(\theta)\ket{\psi_t}$ are both weakly entangled, particularly when the parameters $\theta$ are close to their optimal values.


\section{Circuit optimization: cost function and Trotter-initialisation}\label{app:circ_opt}
As discussed in the main text, the $AQCtensor$ framework involves the design of a parametric quantum circuit with a specified $CNOT$ structure. The parametric circuit is written in terms of so-called $CNOT$ blocks \cite{madden2022best} - defined as a $CNOT$ gate followed by single qubit rotations (see Figure \ref{fig:cnot_block}). Formally, a block with a $CNOT$ gate acting on a ``control" qubit $j$ and ``target" qubit $k$ is written as $\mathrm{CU}_{jk}(\theta_1, \theta_2, \theta_3, \theta_4)$ and for a specified $CNOT$ structure one can then write down a fully parameterised circuit as: 
\beq
\begin{aligned}
V_{\mathrm{ct}}(\boldsymbol{\theta})=& \mathrm{CU}_{\mathrm{ct}(L)}\left(\theta_{3 n+4 L-3}, \ldots, \theta_{3 n+4 L}\right) \cdots \mathrm{CU}_{\mathrm{ct}(1)}\left(\theta_{3 n+1}, \ldots, \theta_{3 n+4}\right) \\
& {\left[R_{z}\left(\theta_{1}\right) R_{y}\left(\theta_{2}\right) R_{z}\left(\theta_{3}\right)\right] \otimes \cdots \otimes\left[R_{z}\left(\theta_{3 n-2}\right) R_{y}\left(\theta_{3 n-1}\right) R_{z}\left(\theta_{3 n}\right)\right] }
\end{aligned}
\eeq
The goal of AQC is to tune the parameters $\theta$ to minimise the cost function under consideration. As discussed in section \ref{sec:algorithm}, when the $CNOT$ structure of the parametric circuit matches that of the second-order Trotter circuit - as is the case for Figures \ref{paramcircuit2} and \ref{fig:trott2} - we can initialise the parameters such that the two circuits are initially identical, and hence such that the state $V(\theta)\ket{0}$ is precisely the Trotter state with time step $dt$ equal to the total simulation time divided by the number of layers $l$ in the circuit Ansatz. This is to allow the the optimization procedure to start from an initial state which is likely to be closer to the target state than a state generated from a random initialisation of $\theta$. More precisely, for a given group of 3 $CNOT$-block repetitions acting on qubits $i$ and $i+1$ - see Figure \ref{paramcircuit2} - we tune the parameters of these blocks such that their combination produces the same 2-qubit unitary as that in Figure \ref{fig:ut} with $\theta = \frac{\pi}{2}-\frac{1}{2}\Delta_i$, $\phi = \frac{1}{2}\alpha_i-\frac{\pi}{2}$ and $\lambda = \frac{\pi}{2}-\frac{1}{2}\beta_i$ where $\alpha_i$, $\beta_i$ and $\Delta_i$ correspond to the coefficients in the Hamiltonian $H_{XYZ}$ in equation (\ref{hxyz}).\\

\noindent We now turn our attention to the cost function in equation (\ref{cost_func_alpha2}). It was pointed out in \cite{khatri2019quantum} that the gradient of the cost function in (\ref{cost_state_prep}) vanishes exponentially in the number of qubits. This observation lead to the distinction between global and local cost functions; local cost functions have only polynomially vanishing gradients in some cases of interest - see \cite{khatri2019quantum, cerezo2021cost, robertson2022escaping} for details. As was shown in \cite{robertson2022escaping}, the Hilbert-Schmidt test - which is a global cost function - can be turned into a local one by adding several ``bit-flip" terms which increases the magnitude of the variance of the gradient:
\beq\label{localcost}
\begin{aligned}
C_{lhs}^{\text{state}} = 1 -  &|\bra{0}V^{\dagger}(\theta)\ket{\psi_t}|^2 - \left(\frac{n-1}{n}\right)\sum\limits_{j=1}^n| \bra{0}X_jV^{\dagger}(\theta)\ket{\psi_t} | ^2\\
-&\left(\frac{n-2}{n}\right)\sum\limits_{j<k}|\bra{0}X_j X_kV^{\dagger}(\theta)\ket{\psi_t} |^2
- ... - \frac{1}{n}\sum\limits_{j<k<l<...}|\bra{0}X_j X_k X_l...V^{\dagger}(\theta)\ket{\psi_t}|^2 
\end{aligned} 
\eeq
One can define a truncated version of $C_{lhs}^{\text{state}}$ to get:
\beq\label{cost_func_alpha}
\begin{aligned}
C_L^{(k)}(\alpha_1,...,\alpha_k) = &1 - |\bra{0}V^{\dagger}(\theta)\ket{\psi_0}|^2 - \alpha_1\sum\limits_{j=1}^n|\bra{0}X_jV^{\dagger}(\theta)\ket{\psi_0}|^2 \\ &-\alpha_2\sum\limits_{j<k}|\bra{0}X_jX_kV^{\dagger}(\theta)\ket{\psi_0}|^2
-...-\alpha_k\sum\limits_{j_1<...<j_k}|\bra{0}X_{j_1}...X_{j_k}V^{\dagger}(\theta)\ket{\psi_0}|^2
\end{aligned}
\eeq
It was shown in \cite{robertson2022escaping}, that when one takes the trivial parametric circuit Ansatz:
\beq\label{ansatz_ex}
V(\theta) = \otimes_{j=1}^{n}e^{-i\theta_j \frac{\sigma_x}{2}}
\eeq
and the trivial target state:
\beq\label{target_ex}
\ket{\psi_0}=\ket{0}
\eeq
the variance of the gradient of $C_L^{(k)}(\alpha_1,...,\alpha_k)$ increases by a constant multiplicative factor for every ``bit-flip" term in equation (\ref{cost_func_alpha}), i.e. we have:
\beq\label{var_clk}
\text{Var}\left[\frac{\partial C_{L}^{(k)}}{\partial \theta_j} \right] \propto \left(\frac{3}{8}\right)^{n-k-1}
\eeq
While this is a rigorous result for the particularly simple circuit Ansatz and target state defined here, we do not have such a proof for the general case. However, the result suggests that the cost function in (\ref{cost_func_alpha}) may converge faster than the cost function in (\ref{cost_state_prep}) for more general cases of interest and we thus use it as a heuristic method - see \cite{robertson2022escaping} for more details on how we tune $\alpha$ throughout the optimization. The cost of calculating $C_L^{(k)}(\alpha_1,...,\alpha_k)$ and its gradient increases with each $k$ so in practice we take $k=1$, i.e. we have dropped all terms with more than one $NOT$ operators $X_i$ in (\ref{cost_func_alpha}), thus leading to the expression in (\ref{cost_func_alpha2}).

\section{Acknowledgements}
This work was funded by the Disruptive Technologies Innovation Fund (DTIF), by Enterprise Ireland, under project number DTIF2019-090 (project QCoIR) and also supported by IBM Quantum.
\newpage

\bibliography{references}{}
\bibliographystyle{unsrt}

\end{document}